\documentstyle[epsfig]{article}

\def\tou#1{{\lower1.2ex\hbox{$\longrightarrow$}\atop
        {\lower-.7ex\hbox{$\scriptscriptstyle #1 $}}}}
\def\lsim{{\lower1.2ex\hbox{$<$}\atop
        {\lower-.7ex\hbox{$\sim$}}}}
\def\gsim{{\lower1.2ex\hbox{$>$}\atop
        {\lower-.7ex\hbox{$\sim$}}}}

\def\be{\begin{equation}}
\def\ee{\end{equation}}

\begin{document}

\begin{titlepage}
\rightline {Si-95-07 \  \  \  \   }

\vspace*{2.truecm}

\centerline{\Large \bf  Universality and Scaling }
\vskip 0.6truecm
\centerline{\Large \bf in Short-time Critical Dynamics }
\vskip 0.6truecm

\vskip 2.0truecm
\centerline{\bf K. Okano$^*$, L. Sch\"ulke, K. Yamagishi$^*$ and B. Zheng}
\vskip 0.2truecm

\vskip 0.2truecm
\centerline{Universit\"at -- GH Siegen, D -- 57068 Siegen, Germany}
\centerline{$^*$ Tokuyama University, Tokuyama-shi, Yamaguchi 754, Japan }

\vskip 2.5truecm

\abstract{ Numerically we simulate the short-time behaviour of the critical
dynamics for the two dimensional Ising model and
Potts model with an initial state of
very high temperature and small magnetization. Critical initial increase 
of the magnetization is observed. The new dynamic critical exponent
$\theta$ as well as the exponents $z$ and $2\beta/\nu$ are
determined from the power law behaviour of the magnetization,
auto-correlation and the second moment. Furthermore the calculation 
has been carried out with both Heat-bath and Metropolis algorithms.
All the results are consistent and therefore universality and scaling
are confirmed.

}

\end{titlepage}

\section{Introduction}

For statistical systems at criticality 
in equilibrium or near equilibrium
it is well known that the critical behaviour is
characterized by universality and scaling.
This is more or less due to the {\it infinite} spatial and time 
correlation lengths. It has long been challenging whether
universality and scaling may also be present for the systems 
far from equilibrium or not. 
An interesting non-equilibrium system 
is the so-called ordering dynamic process:
A statistical system
initially at a high temperature state is suddenly 
quenched to the critical temperature or below where locates the
symmetry broken phase, and then evolves
according to a certain dynamics. 
Domain growth in this process
has been studied since thirty years \cite {bra94}. 
In the case of a system quenched to the neighbourhood
of the critical temperature,
recent investigations show that universality
and scaling may appear in a quite early stage of 
the time evolution where the system is still far from 
equilibrium and the spatial correlation length is small.
Essential here is the large time correlation
length which induces a memory effect.

 Let us consider that the Ising model initially in a random state with
a small magnetization is suddenly quenched to 
 the critical temperature and then evolves
according to a dynamics of model A.  
Janssen, Schaub and Schmittmann \cite{jan89} have
argued by an $\epsilon$-expansion up to two--loop order that,
besides the well known universal  behaviour in the long-time
regime, there exists
another universal stage of the relaxation {\it at early times}, 
the so-called critical initial slip, which sets in right after the
microscopic time scale $t_{mic}$. 
The authors predicted that the magnetization undergoes a {\it critical initial
increase}, and introduced a new dynamic critical exponent
$\theta$ to describe the power law increase
of the magnetization. 
Previously $\theta$  has been measured
with Monte Carlo simulation in two dimensions somehow {\it indirectly}
from the power law decay of the autocorrelation
\cite{hus89,hum91}, and recently in three dimensions
{\it directly} from the power law
increase of the magnetization \cite{li94}.
The exponents $\theta$ for two and three dimensions 
have also been measured from 
damage spreading \cite {gra95}.
They are in good agreement with the result
from the $\epsilon$-expansion.
Detailed scaling analysis reveals 
that the characteristic time scale
for the critical initial slip is
$t_0 \sim m_0 ^ {-z/x_0}$,
where $x_0$ is the scaling dimension of $m_0$, related to $\theta$
by $x_0 =  \theta z + \beta /\nu$.
More interestingly, it was pointed out that the exponents $\beta$,
$\nu$ and $z$ should be valued  the same as those in the
equilibrium or long-time stage of the relaxation.
Therefore, based on the scaling relation in the initial stage
of the time evolution, a new promising way for measuring the
exponents $z$, $\beta$ and $\nu$ has been proposed \cite{li95,li95a,sch95a}.
In reference \cite {li95a,sch95a} the investigation has also
been extended to the critical relaxation starting from
a {\it  completely ordered state }.
 This indicates a possible broad application of the short-time
dynamics  since the universal behaviour of the short-time dynamics
is found to be quite general 
\cite {jan92,oer93,oer94,bra94,rit95,lee95,rut95}. 
More and deeper understanding
of this phenomenon becomes urgent.

Even though analytically the short-time dynamics
has been studied in a variety of systems, most of the numerical 
simulations are limited to the Ising model.
In a previous paper \cite {sch95}, two of the authors have 
systematically extended the numerical investigation of 
the dynamic process discussed above to the critical 
three state Potts model in two dimension with
Monte Carlo Heat-bath algorithm.
They observed the critical initial increase of the magnetization
and have directly measured the critical exponent $\theta$.
The result shows that the direct measurement of $\theta$
is better and more reliable than that indirectly measured from
the power law decay of the auto-correlation, especially when
$\theta$ is smaller.
With $\theta$ in hand, the exponent $z$ can be obtained from 
 the power law decay of the auto-correlation, and then the exponent
$\beta/\nu$ from the power law increase of the second moment
of the magnetization. They are in good agreement with the known
values. Such a procedure relates the direct and indirect measurements
of $\theta$ to each other and 
provides a strong support for the scaling of the 
short-time dynamics for the Potts model. 
On the other hand, it also indicates an efficient way
to estimate the critical exponent $z$ which normally
is quite difficult to be measured due to the critical slowing down.
For the Potts model, among the distributed values of $z$
from different numerical simulations \cite {ayd85,tan87,bin81,ayd88}, 
the result 
confidently supports
the relative small one \cite{ayd85,tan87}.
Compared with the method from the finite size scaling,
 its advantage is that the measurement from the power law behaviour 
is more direct and can be carried out
in a single lattice.

The first purpose of this paper is to present a detailed description
of the results for the critical Potts model reported briefly in the
previous paper \cite{sch95}. The results have been refined by taking the
effect of $t_{mic}$ into account, whose importance will be
clarified in the
investigation of the universality described below. Furthermore the
data have been extended to bigger lattice $L=576$ for the simulation
of $m_0=0.0$ and smaller initial magnetization $m_0=0.02$ for the
measurement of $\theta$ in order to clarify some unclear
points.

The second purpose of this paper is to investigate the
universality for short-time dynamics. We perform all of the simulations
both with Heat-bath and Metropolis algorithms. 
We numerically observe the microscopic time
scale $t_{mic}$, after which the universality will be switched on.
Depending on what kind of dynamics one takes,
$t_{mic}$ is different. Such a knowledge becomes very important 
in the measurement of the universal
critical exponents. In this paper we have also
presented a systematic investigation of the two dimensional 
Ising model (two state Potts model).
Even though
there exist already several simulations for the
Ising model, up to now no serious confirmation of the universality was
presented in the sense of numerical simulation, especially for the
direct measurement of $\theta$.  In reference \cite {men94}, the auto-correlation
for the Ising model was calculated with both Heat-bath and Metropolis
algorithms. However the results were rather rough.
Our results complete the systematic simulation of the 
short-time dynamics for the Ising model with better statistics.

In section 2, we will briefly recapitulate the scaling behaviour 
for the short-time
dynamics, which will serve as the theoretical base of 
the numerical simulation.
In section 3, we will describe the result of a simulation for the Ising model
in two dimensions, concentrating our attention to the universality. 
In section 4, the results for the three state Potts model will be given.
Conclusion and some remarks are given in section 5.

\section{Scaling for the short-time dynamics}

Let us consider a dynamic system of model A.
Janssen, Schaub and Schmittmann have shown \cite{jan89} that 
when a system initially in a state with very high
temperature $T \gg T_c$ is
 suddenly quenched to the
critical temperature and then evolves according to
a certain dynamics, besides the well-known universal
behaviour in the long-time regime, there emerges another 
universal stage of the dynamic relaxation at the {\it macroscopic
short-time regime}, which sets in right 
after a microscopic time scale $t_{mic}$.
For the $O(N)$ vector model the renormalization of the 
Langevin dynamics with initial conditions has been
investigated with $\epsilon$-expansion up to two loop.
An interesting observation is that a new divergence is
induced in the short-time dynamics which should be 
renormalized by the initial magnetization.
Taking this point into account, a generalized dynamic scaling
relation has been derived, 
\begin{equation}
M^{(k)}(t,\tau,m_{0})=b^{-k\beta/\nu}
M^{(k)}(b^{-z}t,b^{1/\nu}\tau,
b^{x_{0}}m_{0}) ,
\label{etwo2}
\end{equation}
where $M^{(k)}$ is $k-$th moment of the magnetization,
$t$ is the dynamic relaxation time,
$\tau$ is the reduced temperature,
the parameter $b$ represents the spatial rescaling factor
 and in addition $x_0$ is 
the anomalous dimension
of the initial magnetization $m_0$. It is shown 
that $x_0$ is {\it a new independent exponent}, i.e. it can not
be expressed by other known critical exponents.

As an example, let us now consider the time evolution
of the magnetization in the initial stage of the dynamic relaxation.
From above scaling relation (\ref{etwo2}), taking 
$\tau=0$
and $b=t^{1/z}$, 
\begin{equation}
M(t,m_{0})=t^{-\beta/\nu z}
M(1,t^{x_{0}/z}m_{0}).
\label{etwo3}
\end{equation}
In our notation $M \equiv M^{(1)}$
and the argument $\tau$ has been omitted.
Assuming the initial magnetization $m_{0}$ is small, 
the time evolution of
the magnetization may be expanded according to $m_0$
\begin{equation}
M(t,m_{0})=m_{0} F(t) + O(m_0^2).
\label{etwo4}
\end{equation}
Here the condition $M(t,m_{0}=0)\equiv0$ has been used.
From equations (\ref{etwo3}) and 
(\ref{etwo4})
one can easily realize that the time evolution of 
the magnetization obeys a power law 
\begin{equation}
M(t) \sim m_0 \, t^\theta
\label{etwo1}
\end{equation}
where the exponent $\theta$ which we name `short-time dynamic
exponent', is related to $x_0$ by
\begin{equation}
 \theta =(x_0 - \beta /\nu)/z.
\label{etwoextra}
\end{equation}
Here we should stress that the  power law behaviour
is valid only in case that $t^{x_{0}/z}m_{0}$ is also
small enough. Therefore the universal behaviour shown in
(\ref{etwo1}) is expected in the initial stage of the time 
evolution. The time scale for it is $t_0 \sim m_0 ^ {-z/x_0}$.
However, in the limit of $m_0=0$ the time scale
$t_0$ goes to infinity. Hence the initial condition can leave
its trace even in the long-time regime \cite {die93,rit95,rit95a}. 
Interestingly, for the $O(N)$ vector model
($n=1$ corresponds to the Ising model) it is shown by
$\epsilon$-expansion that $x_0 > \beta / \nu$ and therefore
$\theta > 0$, i.e. the magnetization really undergoes
an {\it initial increase}. This has also been confirmed 
by the numerical simulation for the three dimensional
 Ising model directly
and also by the study of the damage spreading \cite {gra95}.

As the spatial correlation length in the beginning
of the time evolution is small, for a finite system
of dimension d
with lattice size $L$ the second moment $M^{(2)}(t,L) \sim
L^{-d}$. From the finite size scaling one can deduce 
\begin{equation}
M^{(2)}(t) \sim t ^ {(d-2\beta / \nu)/z}.
\label{efou2}
\end{equation}
Furthermore careful scaling analysis shows that auto-correlation
also decays by a power law \cite {jan92}
\begin{equation}
A(t) \sim t^{-d/z+\theta}.
\label{ethr2}
\end{equation}
The new short-time dynamic exponent
$\theta$ enters the auto-correlation.
Actually the first numerical estimation of $\theta$
is from the measurement of the exponent $\theta - d/z$
\cite {hus89,hum91}.
Taking the exponent $z$ as an input, one obtains
$\theta$. However, usually $z$ is not known so accurately.
Since $\theta$ is normally much smaller than $z$ as well as
$-d/z+\theta$, a small relative error of $z$ and $-d/z+\theta$
may induce a big error for $\theta$. 

Our strategy is that we first measure the exponent $\theta$
directly from the power law increase of the magnetization,
then taking it as an input we estimate the exponent $z$
from the auto-correlation, and with $z$ in hand we finally 
obtain the static
exponent $2 \beta/\nu$ from the second moment.
Such a procedure can provide strong confirmation for
the scaling relation for the short-time dynamics.

Traditionally the exponent $z$ is defined in the long-time regime
of the dynamic process and normally measured from the
exponential decay of the auto-correlation or the magnetization
of the systems. This measurement is difficult due
to the critical slowing down. However, if we can obtain
$\theta$ from the direct measurement of the initial increase of
the magnetization (\ref{etwo1}), $z$ obtained
from the scaling behaviour of the auto-correlation in (\ref{ethr2})
can be quite rigorous.
 One may
also expect that the measurement is to some extent free from the
critical slowing down, since all of these quantities are measured in
the short-time regime of the dynamic process. 

\section{The Ising model}

The Hamiltonian for the Ising model is
\begin{equation}
H=J  \sum_{<ij>}  S_i\ S_j\;,\qquad S_i=\pm 1\;,
\label{hamii}
\end{equation}
with $<ij>$ representing nearest neighbours.
In the equilibrium the Ising model is exactly solvable.
The critical point locates at $J_c=\log(1+\surd 2)/2$.
In principle any type of the dynamics can be given to the system
to study the non-equilibrium evolution processes.
In this paper we concentrate our attention on the 
Monte Carlo Heat-bath and Metropolis algorithm,
both of which belong to the dynamics of model A.

\subsection{Magnetization}

As discussed in section 2, we study the short-time
behaviour of the dynamic process starting
from an initial state with {\it zero correlation length and
small magnetization}. 
Such initial configurations can easily be generated
numerically.
 Starting from those initial
configurations, the system is updated both with the Heat-bath 
and Metropolis
algorithm, in order to confirm the universality. 
We measure the time evolution of the magnetization
\begin{equation}
M(t)= \frac{1}{N}\,\langle\sum_i S_i(t)\rangle.
\end{equation}
where $N$ is the number of the lattice sites and the average
$< \cdots >$ is taken over the independent initial 
configurations and the random force. The total number of the independent initial 
configurations is $150\ 000$ for $m_0=0.08, 0.06$ and $0.04$ and
$300\ 000$ for $m_0=0.02$. Errors are estimated by dividing
the data into five groups.

In Fig. 1a and b the time evolution of the magnetization
in double log scale with $m_0=0.02$ for different lattice sizes
for both the Heat-bath and Metropolis algorithm, respectively, is
displayed. For the Heat-bath algorithm, one can clearly see the
initial {\it power law} increase of the magnetization 
from a very early stage of
the time evolution, i.e. the microscopic time scale 
$t_{mic}$ is ignorably small.
 On the other hand, for the Metropolis algorithm,
this is not the case in the very beginning of the time
evolution. 
The power law increase of the magnetization becomes stable
only  after certain
time steps, say $t\sim 20$ to $30$.
In other words, for the Metropolis algorithm $t_{mic}\sim 20$ to $30$,  
which is bigger than that for Heat-bath algorithm.
 In order to see this more clearly, we plot 
in Fig. 2 the exponent $\theta$ as a function of 
the time $t$ for both Heat-bath and
Metropolis algorithms for lattice size $L=128$ and different
initial magnetization $m_0$. Here,
$\theta$ at time $t$ is measured from the slope of the
curves by the least square fit in the time interval of $[t,t+15]$.

As expected, the exponent $\theta$ for the
Heat-bath algorithm is quite stable from the very beginning of
the time evolution but not that for Metropolis.
Taking into account the errors as well as
the fluctuation in time direction, however, 
the exponents $\theta$ from both algorithms
 become the same after some time steps around $t \sim 30$. 
This is a real indication of the universality
in the short-time dynamics.
From the above procedure  we may summarize a criterion
 to measure $\theta$ and also other exponents
discussed later:

(i) We first scan the data by the exponent measured at
each time $t$ by least square fitting 
in the time interval of $[t,t+15]$. We call this `$15$-scan'
in the following. Of course, the number of data
for the least square fit can differ from 15 which is used here.

(ii) Using the figure obtained from the $15$-scan, we can estimate
$t_{mic}$ from which the exponent becomes stable.
If we perform a simulation with different algorithms,
we can compare these results and see that the universality 
is switched on after the microscopic
time $t_{mic}$.

(iii) Finally we perform the least square fit in 
the time interval of $[t_{mic},T]$ to obtain the final values for 
the exponents.
Here $T$ can normally be the maximum updation time
where finite size effects and the finite $m_0$ effect have
not shown up. But sometimes we may take a bit smaller $T$ 
in order to escape the big fluctuation due to the lack of
statistics. This can be judged by an inspection of the result of
the 15-scan. 

Coming back to Fig.2,
one may observe a slight tendency that $t_{mic}$ decreases
as $m_0$ gets smaller.
We should stress that the errors estimated here
can not completely represent the fluctuations in the time direction
due to the large time correlation length and also
other systematic errors, e.g. those from the random numbers.
In table 1, results for $\theta$
measured from a time interval $[30,100]$ are listed. 
A detailed analysis of the data reveals that the finite size effect 
is quite small for a lattice size $L=128$.

From $m_0=0.08$ down to $m_0=0.02$ the measured $\theta$
shows a smooth linear increase. By definition $\theta$ should be
measured in the limit of $m_0=0$. 
Following the procedure in the previous paper \cite {sch95}
we have carried out a linear extrapolation to $m_0=0$ and listed
the results also in the table 1.
The value $\theta=0.191(1)$ from the Heat-bath algorithm is
well consistent with the value $\theta=0.191(3)$  obtained from
damage spreading \cite {gra95}
and those obtained from auto-correlation before \cite {hus89,hum91}.
For Metropolis algorithm, our first direct measurement 
$\theta=0.197(1)$ is very close to that for
Heat-bath and gives strong support for universality.
The slight difference of $\theta$ for Heat Bath 
and Metropolis algorithms may come from the remaining of the
finite size effect, 
finite $m_0$ effects or other systematic errors.
To check this point
bigger simulation with high statistics for smaller
$m_0$ and bigger lattice size or
even for longer updation time may be needed.

\begin{table}[h]\centering
$$
\begin{array}{|c|lllll|}
\hline
 m_0  &\quad 0.08 &\quad 0.06 &\quad 0.04 &\quad 0.02 &\quad 0.00 \\
\hline
 HeatB & 0.173(1) &  0.179(1)& 0.183(1) & 0.187(1) & 0.191(1) \\
\hline
 MetroP & 0.173(1) & 0.182(1)  & 0.187(1)  &  0.192(1) & 0.197(1)\\
\hline
\end{array}
$$
\caption{ The short-time dynamic exponent $\theta$ measured
for lattice size $L=128$ with different initial magnetization
for the Ising model.  }
\label{T1}
\end{table}

\subsection{Auto-correlation}

Now we set $m_0=0$. The auto-correlation is defined as
\begin{equation}
\label{ethr1_is}
A(t)=\frac{1}{N}\, \langle\sum_i S_i(0)S_i(t)\rangle,
\end{equation}
We have performed the simulation with both the Heat Bath 
and Metropolis algorithm for lattice sizes $L=256$.    
The number of independent initial configurations
for the average is $35\ 000$.

In order to see the universality and
the possible effect of $t_{mic}$ as well as the fluctuation
in the time direction, we again perform the 15-scan and
display in Fig. 3 the exponent 
$-d/z+\theta$ as a function of time $t$ for both Heat-bath
and Metropolis algorithm. 

After a microscopic time scale
$t_{mic} \sim 30$, the results from both algorithms 
agree well and are presenting a quite stable
power law behaviour. This  again supports
the universality.  Within the errors both algorithms give
almost the same results. However, the error for
Metropolis is much bigger than that for the Heat-bath algorithm.

In table 2 the exponents $-d/z+\theta$ measured from the time interval
$[30,100]$ are listed. They are consistent with
the previous measurement \cite {men94}, but the errors are much smaller.
The dynamical exponent $z$ obtained by taking
 $\theta$ measured in the
previous subsection as input are also given in table 2.
For the Heat-bath algorithm, the value $z=2.155(3)$ is 
in good agreement with $z=2.153(2)$ measured from the
finite size scaling of the Binder cumulant \cite {li95a}.
The value $z=2.137(11)$ for Metropolis algorithm
is slightly smaller but roughly consistent within the errors.
Actually in reference \cite {li95a}, depending on the observables 
and the dynamic processes used for estimating $z$,
 the values for $z$ from the
finite size scaling are also varying within $1 \%$.
To get more accurate $z$ still requires high precision
numerical measurement.

\begin{table}[h]\centering
$$
\begin{array}{|c|l|l|l|l|}
\hline
   & \theta - d/z &\quad z & (d-2 \beta/\nu)/z & \ \ 2 \beta/\nu\\
\hline
 HeatB & 0.737(1) &  2.155(03)& \ \ 0.817(7) & 0.240(15)\\
\hline
 MetroP & 0.739(5) & 2.137(11)  & \ \  0.819(5) & 0.250(14)\\
\hline
\end{array}
$$
\caption{ The exponents measured
for lattice size $L=256$ with initial magnetization $m_0=0.0$
for the Ising model.  }
\label{T2}
\end{table}

\subsection{Second moment}

For $m_0=0$, the second moment is defined by 
\begin{equation}
M^{(2)}(t)= \frac{1}{N^2} \,\langle\left[ \sum_i S_i(t)\right] ^2\rangle.
\label{efou1_is}
\end{equation}
For the Heat-bath algorithm the exponent $(d-2\beta / \nu)/z$
 is quite stable after
$t_{mic}\sim 20$ to $30$. However,  for the Metropolis algorithm,
$t_{mic}$ seems to be somewhat bigger, $t_{mic} \sim 60$. In
order to obtain more reliable results, 
we have extended the number of time steps
for the Metropolis algorithm up to 150.
 In table 2 the measured values for $(d-2\beta / \nu)/z$
together with the exponent $2 \beta / \nu$ deduced by taking
the value of $z$ from the previous subsection as input are given.
All the results are consistent and confirm universality.

We should point out that the determination of the 
exponent $2\beta / \nu$ is not very accurate here  since
the exponent $2\beta / \nu$ is much smaller than
$d$ and $z$, and therefore small relative errors in
$z$ and $(d-2\beta / \nu)/z$ will induce big errors
for $2\beta / \nu$.

\vskip 0.5 truecm

In this section we have investigated the universality
in short-time dynamics for the Ising model. We have numerically observed
the existence of the microscopic time scale $t_{mic}$, after which
the universality will be switched on. Depending on the algorithm,
Metropolis or Heat-bath,
$t_{mic}$ is different. Such a knowledge is very
important in the  measurement of the critical exponents
 in short-time dynamics.
In the next section we perform a detailed numerical simulation for the 
Potts model, taking this knowledge into account carefully.

\section{The Potts model}

The Hamiltonian for the $q$ state Potts model is given by
\begin{equation}
H=J  \sum_{<ij>}  \delta_{\sigma_i,\sigma_j},\qquad \sigma_i=1,...,q
\label{hamip}
\end{equation}
where $<ij>$ represents nearest neighbours. It is known that the critical
points locate at $J_c=\log(1+\sqrt{q})$.The Ising model
is the two state
($q=2$) Potts model. In this section, we investigate 
the three state ($q=3$) Potts model
 in two dimensions. 

\subsection{Magnetization}

We measure the time evolution of the magnetization defined as
\begin{equation}
M(t)= \frac{3}{2 N}\,\langle\sum_i 
      \left(\delta_{\sigma_i(t),1}-\frac{1}{3}\right)\rangle.
\end{equation}
The total number of the independent initial 
configurations used for taking the average is $80\ 000$
for $m_0=0.06$ and $0.08$ and $600\ 000$ 
for $m_0=0.02$ and $0.04$.
Similar to the case of the Ising model,
errors are estimated by dividing
the data into two or four groups.
  
In Fig. 4 the time evolution of the magnetization
in double log scale 
with initial value $m_0=0.04$ for different lattices
and for both the Heat-bath and Metropolis algorithm is displayed.
Somewhat different from the case of the Ising model,
the outlooks of the curves from the
Heat-bath and Metropolis algorithms appear very different.
For the curves from Heat-bath algorithm the power law behaviour 
starts right at the very beginning of the time evolution.
This means the microscopic time scale  $t_{mic}$ is small and ignorable.
However for the Metropolis algorithm they first decrease and then
increase after some time steps. Later analyses show that 
the power law behaviour becomes stable only after
around $20$ updation time steps, i.e. $t_{mic} \sim 20$.

In the previous paper \cite {sch95}, the simulation was
carried out using only the Heat-bath algorithm. Due to the fact
that the error at the beginning of the time evolution is the smallest,
we simply measured the exponent from the first $15$ time steps.
However, the result of the simulation for the Metropolis algorithm
indicates that we should carefully analyse
the data, with special attention to the effect of $t_{mic}$.
We show in Fig. 5 the exponent $\theta$ vs. $t$, obtained by
the $15$-scan for both the Heat-bath and Metropolis algorithm
with initial magnetization $m_0=0.04$. 
It is clear that $\theta$ from the Heat-bath algorithm
is quite stable from the very beginning of the time evolution
but for the Metropolis this is apparently not the case.
However, as was the case of the Ising model, the exponent $\theta$ from 
both the Heat-bath and Metropolis algorithm
coincide after some time steps around $t_{mic}\sim 20$, showing again
universality in short-time dynamics.

In table 3 the values of $\theta$
measured from the time interval $[20,100]$ are listed.
In principle for the Heat-bath algorithm the measurement may
be carried out from the beginning. However for the reason 
of comparison we like to treat both algorithms the same.
If we compare the results for the Heat-bath algorithm obtained
in this paper with those measured from the first fifteen time steps
in the previous paper \cite {sch95}, they are quite near.

In order to see the finite size effect, we have plotted in
Fig. 6 the results of lattices $L=72$ and $L=144$
with the initial magnetization $m_0=0.02$ for the Heat-bath algorithm.
Within the errors they are overlapping. 
Therefore we are satisfied with the lattice
size $L=72$ for the measurement of $\theta$.
In table 3 the averaged values of $\theta$ 
for both algorithms still show some slight difference
even though for smaller $m_0$ it looks as if they 
can be covered by the errors. Similar to the case of the 
Ising model this may be the remnant
of the finite size or finite $m_0$ effects, or other
systematic errors. 

From $m_0=0.08$ down to $m_0=0.02$ the measured $\theta$
shows also a smooth linear decrease.
Therefore
we carry out a linear extrapolation for $\theta$
to the initial magnetization $m_0=0$.
Since here we have measured $\theta$ in a different 
time regime, the result given in the previous
paper \cite {sch95} for $\theta$ has slightly been modified.
Due to the extra data for $m_0=0.02$ the result
extrapolated to $m_0=0.0$ is more reliable.

\begin{table}[h]\centering
$$
\begin{array}{|c|lllll|}
\hline
 m_0  &\quad 0.08 &\quad 0.06 &\quad 0.04 &\quad 0.02 &\quad 0.00 \\
\hline
 HeatB & 0.110(1) &  0.100(1)& 0.092(2) & 0.084(3) & 0.075(3) \\
\hline
 MetroP & 0.100(1) & 0.092(1)  & 0.084(1)  &  0.077(2) & 0.070(2)\\
\hline
\end{array}
$$
\caption{ The short-time dynamic exponent $\theta$ measured
for lattice size $L=72$ with different initial magnetization
fir the Potts model.  }
\label{T3}
\end{table}

\subsection{Auto-correlation}

The auto-correlation for the Potts model is given by
\begin{equation}
\label{ethr1}
A(t)=\frac{1}{N}\, \langle\sum_i 
      \left(\delta_{\sigma_i(0),\sigma_i(t)}-\frac{1}{3}\right)\rangle.
\end{equation}

In Fig. 7 the auto-correlation for different lattice sizes for
both algorithms is displayed in double log scale. 
Although in case of the magnetization already for a lattice size $L=72$
a nice power law increase is already observed,
 for the auto-correlation the lattice size
$L=72$ is not big enough to present the power law 
behaviour. The convergence to
a power law decay only starts around  a lattice size $L=144$.
This can especially be seen for the Heat-bath algorithm.
Compared with the Heat-bath algorithm the results from
the Metropolis algorithm fluctuate a bit depending on the
size of the lattice.
As in the previous section, in order to show the universality and
the possible effect of $t_{mic}$ as well as the fluctuation
in the time direction, we present the exponent 
$-d/z+\theta$ obtained by the 15-scan for $L=288$ as a function 
of time $t$ for both the Heat-bath and Metropolis
algorithm in Fig. 8.

For the lattices bigger than $L=144$,
after a microscopic time scale $t_{mic} \sim 5$ the curves for
both algorithms coincide and are presenting quite stable
power law behaviour even though the fluctuations after
time $t \sim 50$ become very big. Such a small  $t_{mic}$
here is consistent with the scenario in the last section that the
smaller initial magnetization the shorter $t_{mic}$ is.
Within the errors both algorithms give almost the same results.
However we should point out that the error for Metropolis is much bigger
than that for the Heat-bath algorithm.
All these show that for the study of short-time dynamics
Heat-bath algorithm is more efficient.
In table 4 the exponent $-d/z+\theta$ measured from the time interval
$[5,50]$ is listed. To avoid too big fluctuations
we have not made the measurement up to $t=100$.
In the previous paper \cite {sch95} from lattice
size $L=144$ and $L=288$, a linear extrapolation to
infinite lattice size was carried out. However, the result for
lattice size $L=576$ does not go in this direction.
Actually the difference among the results for lattice
$L=144,288,576$ is already very small as it was also pointed out
in the previous paper. Therefore in this paper
the result for infinite lattice
is given as a simple average of the three lattices.
The situation for Metropolis is less satisfactory.
Anyway we also give the same average over the three lattices.
From these values as well as those for $\theta$ in the previous
subsection we can obtain the exponent $z$.

One can now realize what was mentioned in section 2, i.e., 
from the measurement
of the auto-correlation a quite rigorous 
value for $z$ can be obtained in case
that $\theta$ is known. Compared to the values of $z$ distributed
between $z=2.2$ and $z=2.7$ from different numerical measurement 
\cite {ayd85,tan87,bin81,ayd88},
 our result supports the relative small $z$ \cite{ayd85,tan87}.
The results for both algorithms
coincide very well.

\begin{table}[h]\centering
$$
\begin{array}{|c|llll||l|}
\hline
 L   &\quad 144 &\quad 288 &\quad 576 &\quad \infty &\quad z \\
\hline
 HeatB &  0.839(1)& 0.834(1) & 0.835(1) & 0.836(2) & 2.196(08) \\
\hline
 MetroP & 0.849(3) & 0.831(2) & 0.843(6) & 0.841(5) & 2.198(13)\\
\hline
\end{array}
$$
\caption{ The exponent $\theta-d/z$ measured
for different lattice sizes with initial magnetization $m_0=0.0$
for the Potts model. The last column gives the values for $z$. }
\label{T4}
\end{table}

\begin{table}[h]\centering
$$
\begin{array}{|c|lll||l|}
\hline
 L  &\quad 144 &\quad 288  &\quad \infty &\quad  2 \beta/\nu\\
\hline
 HeatB  &  0.789(2)& 0.787(2) & 0.788(1)  & 0.269(07) \\
\hline
 MetroP  & 0.787(7) & 0.789(9) & 0.788(6) & 0.269(16)\\
\hline
\end{array}
$$
\caption{ The exponent $(d-2 \beta/\nu)/z$ measured
for different lattice sizes with initial magnetization $m_0=0.0$
for the Potts model. The last column gives the values for
the exponent $2 \beta/\nu$. }
\label{T5}
\end{table}

\subsection{The second moment}

The second moment for the Potts model is defined by 
\begin{equation}
M^{(2)}(t)= \frac{9}{4 N^2 q} \sum_q \,\langle\left[ \sum_i 
    (\delta_{\sigma_i(t),1} -\frac{1}{3})\right] ^2\rangle.
\label{efou1}
\end{equation}
Here we have taken the average of different components q
of the variable $\sigma_i(t)$. (In the simulation of 
the magnetization $M(t)$ starting from
a non-zero initial magnetization, in principle we may also measure
the exponent $\theta$ from the power law behaviour  of 
other components. However we did not do since the quality of the
power law behaviour for different components is not exactly the same.)

In Fig. 9 the time evolution of the second moment for
different lattices for both algorithms is plotted. 
We again
plot the exponent $(d-2\beta / \nu)/z$ for each
time step $t$ in Fig. 10. It is clear that $t_{mic}$
for Heat-bath is shorter than that for Metropolis.
Results for both algorithms coincide after around
$t_{mic} \sim 30$ which is also bigger than $t_{mic}$ for
$\theta$ and $-d/z+\theta$.
In table 5 the measured values for $(d-2\beta / \nu)/z$
are given. For the Heat-bath algorithm
one may start the measurement from $t \sim 10$. However
for the reason of comparison we treat both algorithms the same 
and perform measurements within the time interval 
$[35,100]$. The result for the lattice $L=576$ 
is a bit fluctuating and is not given in table 5.
 Since the finite size effect 
is already smaller than the statistical fluctuation,
the value of $(d-2\beta / \nu)/z$ for infinite lattice
is given as a simple average over $L=144, 288$.

\section{Conclusions}

We have simulated the universal short-time behaviour of
critical dynamics for the Ising model and the Potts model with
an initial state of very high temperature and small magnetization.
The critical initial increase of the magnetization 
is observed.  
From the power law behaviour
of the magnetization $M(t)$, the second moment $M^{(2)}(t)$
and the auto-correlation $A(t)$, we obtain the related critical exponents
$\theta$, $z$ and $\beta/\nu$ and confidently confirm
the scaling properties in the short-time
dynamics. The direct measurement of $\theta$ 
shows the advantage of the method. 
Especially it allows a rigorous determination
of the exponent $z$, which is normally quite difficult
to be measured from the long-time regime
of the dynamic process. Furthermore all the simulations are carried out
with both the Heat-bath and Metropolis algorithm. The results are consistent
and the universality in the short-time dynamics is confirmed

Here we should mention that not all the models would
have a positive $\theta$. For example, for the  Potts model
with $q=4$ the exponent $\theta$ is likely negative or very close to zero.
In this case the measurement of $\theta$ will become 
more difficult. On the other hand, how to determine the
exponent $\nu$ as well as the critical point from the power law behaviour
of the observables in
the short-time dynamics is also very interesting. 

Finally we would like to point out that the investigation
of the short-time dynamics
for statistical systems may be extended to the
{\it dynamic} field theory, e.g. the stochastically 
quantized field theory where a {\it fictitious} 
dynamic process is introduced and the conventional field
theory is approached in the equilibrium \cite {par81,nam92}. 
Detailed investigations have been performed, especially for gauge
theory and complex systems \cite {oka93a,fuj94}. 
However, up to now all these studies
are only concentrated to the long-time behaviour 
of the dynamic process and its equilibrium.
It will be very interesting to know whether
the properties of the conventional field
theory may be already obtained from the short-time
behaviour of the dynamic system or not.
Such a knowledge will be important for the numerical simulation
of the lattice gauge theory.

Acknowledgement: L.S. and B.Z would like to thank Z.B. Li
for helpful discussions and K. Untch for maintaining the
work stations. K.O. is grateful to the DAAD and the JSPS
for financial support
for his stay in Germany, during which part of the calculations
was carried out.

%\bibliographystyle{pr_NP}
%\bibliography{//ising}

\newpage
\addtolength{\topmargin}{-5cm}
\addtolength{\textheight}{5cm}

\renewcommand{\thefigure}{1.a}

\begin{figure}[t]\centering 
\epsfysize=12cm
\epsfclipoff
\fboxsep=0pt
\setlength{\unitlength}{1cm}
\begin{picture}(13.6,12)(0,0)
\put( 0.1, 7.7){\makebox(0,0){$\frac{M(t)}{M(0)}$}}
\put( 9.8,  .2){\makebox(0,0){$t$}}
\put( 6.5, 3.0){\makebox(0,0){\bf\large Heat-bath}}
\put(10.0, 6.0){\makebox(0,0){L=16}}
\put( 7.8, 6.5){\makebox(0,0){32}}
\put(10.5, 6.8){\makebox(0,0){64,128}}
\put(0,0){{\epsffile{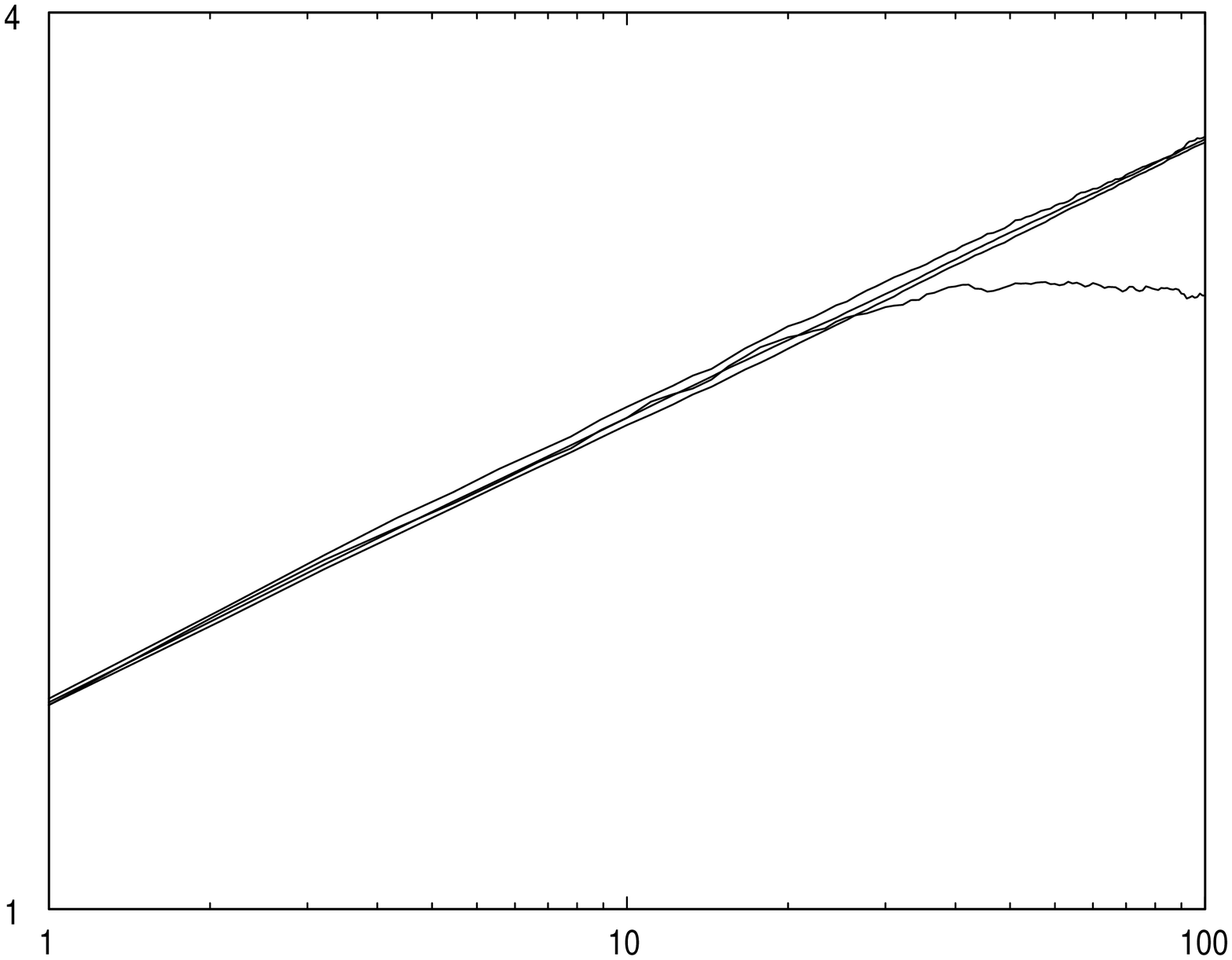}}}
\end{picture}
\caption{
Time evolution of the magnetization in double log scale for the
Ising model with $m_0$=0.02 for the Heat-bath algorithm.
}
\label{fig1a}
\end{figure}

\renewcommand{\thefigure}{1.b}

\begin{figure}[b]\centering 
\epsfysize=12cm
\epsfclipoff
\fboxsep=0pt
\setlength{\unitlength}{1cm}
\begin{picture}(13.6,12)(0,0)
\put( 0.1, 7.7){\makebox(0,0){$\frac{M(t)}{M(0)}$}}
\put( 9.8,  .2){\makebox(0,0){$t$}}
\put( 6.5, 3.0){\makebox(0,0){\bf\large Metropolis}}
\put( 9.0, 4.5){\makebox(0,0){L=16}}
\put(10.0, 6.5){\makebox(0,0){32}}
\put( 8.3, 7.1){\makebox(0,0){64,128}}
\put(0,0){{\epsffile{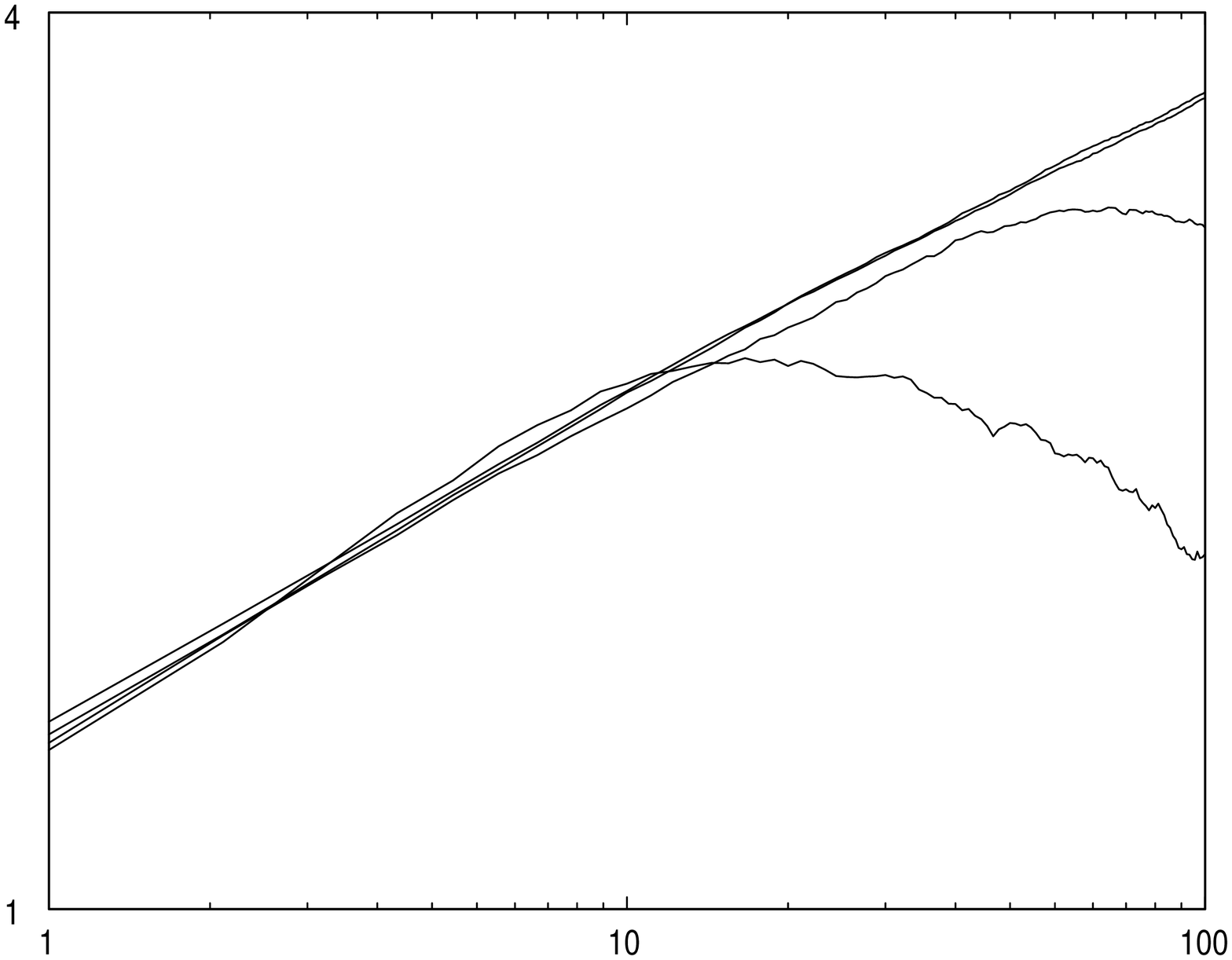}}}
\end{picture}
\caption{
Time evolution of the magnetization in double log scale for the
Ising model with $m_0$=0.02 for the Metropolis algorithm.
}
\label{fig1b}
\end{figure}

\renewcommand{\thefigure}{2.a}

\begin{figure}[t]\centering 
\epsfysize=12cm
\epsfclipoff
\fboxsep=0pt
\setlength{\unitlength}{1cm}
\begin{picture}(13.6,12)(0,0)
\put( 0.7, 8.1){\makebox(0,0){$\theta$}}
\put( 9.8,  .2){\makebox(0,0){$t$}}
\put( 6.5, 8.0){\makebox(0,0){\bf\large $\times$\ \ Metropolis}}
\put( 6.5, 2.5){\makebox(0,0){\bf\large $\Box$\ \ Heat-bath, $m_0$ = 0.02,
   L=128}}
\put(0,0){{\epsffile{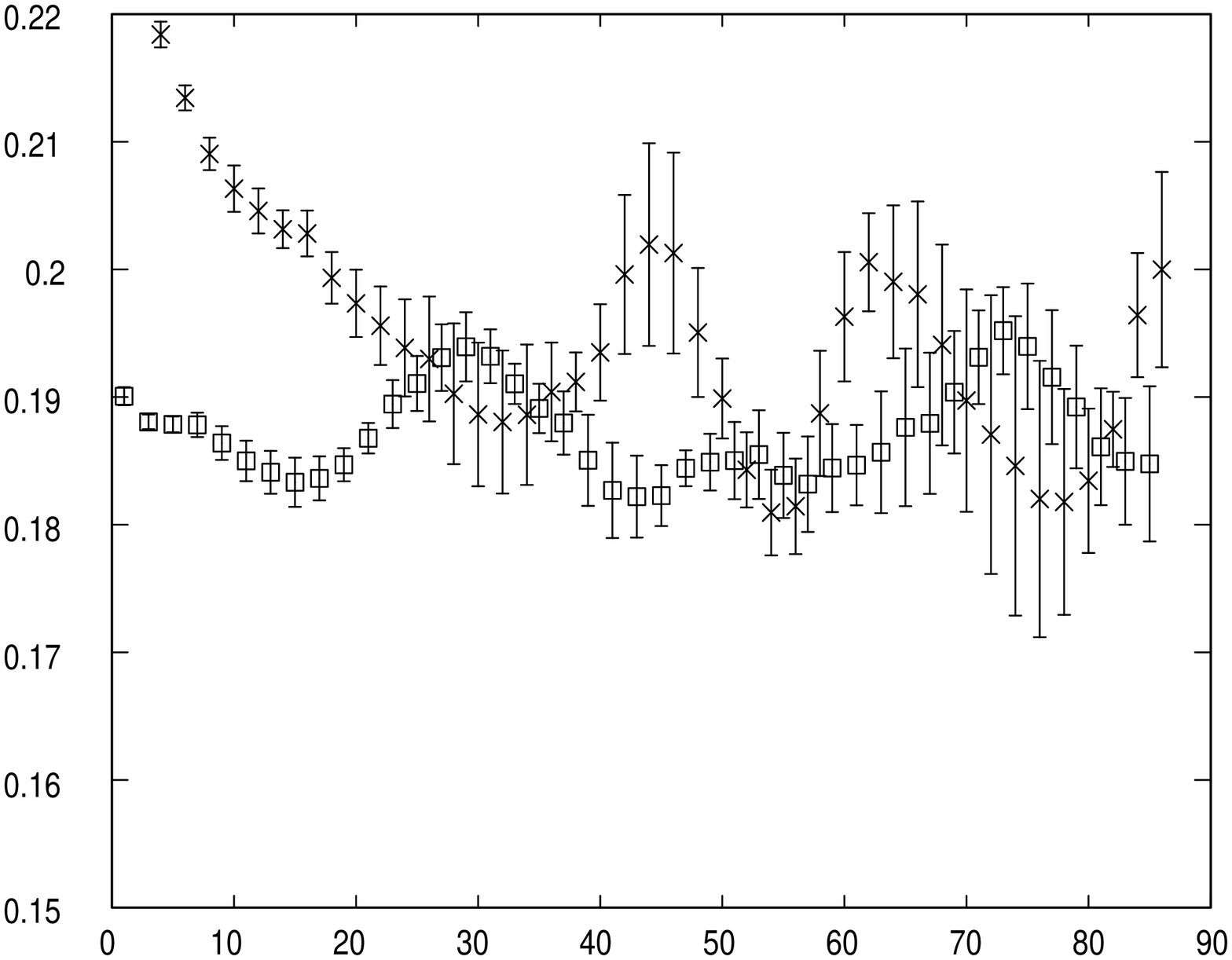}}}
\end{picture}
\caption{
$\theta$ vs. $t$ for the Ising model with $m_0$=0.02.}
\label{fig2a}
\end{figure}

\renewcommand{\thefigure}{2.b}

\begin{figure}[b]\centering 
\epsfysize=12cm
\epsfclipoff
\fboxsep=0pt
\setlength{\unitlength}{1cm}
\begin{picture}(13.6,12)(0,0)
\put( 0.7, 8.1){\makebox(0,0){$\theta$}}
\put( 9.8,  .2){\makebox(0,0){$t$}}
\put( 6.5, 7.0){\makebox(0,0){\bf\large $\times$\ \ Metropolis}}
\put( 6.5, 2.5){\makebox(0,0){\bf\large $\Box$\ \ Heat-bath, $m_0$ = 0.04,
   L=128}}
\put(0,0){{\epsffile{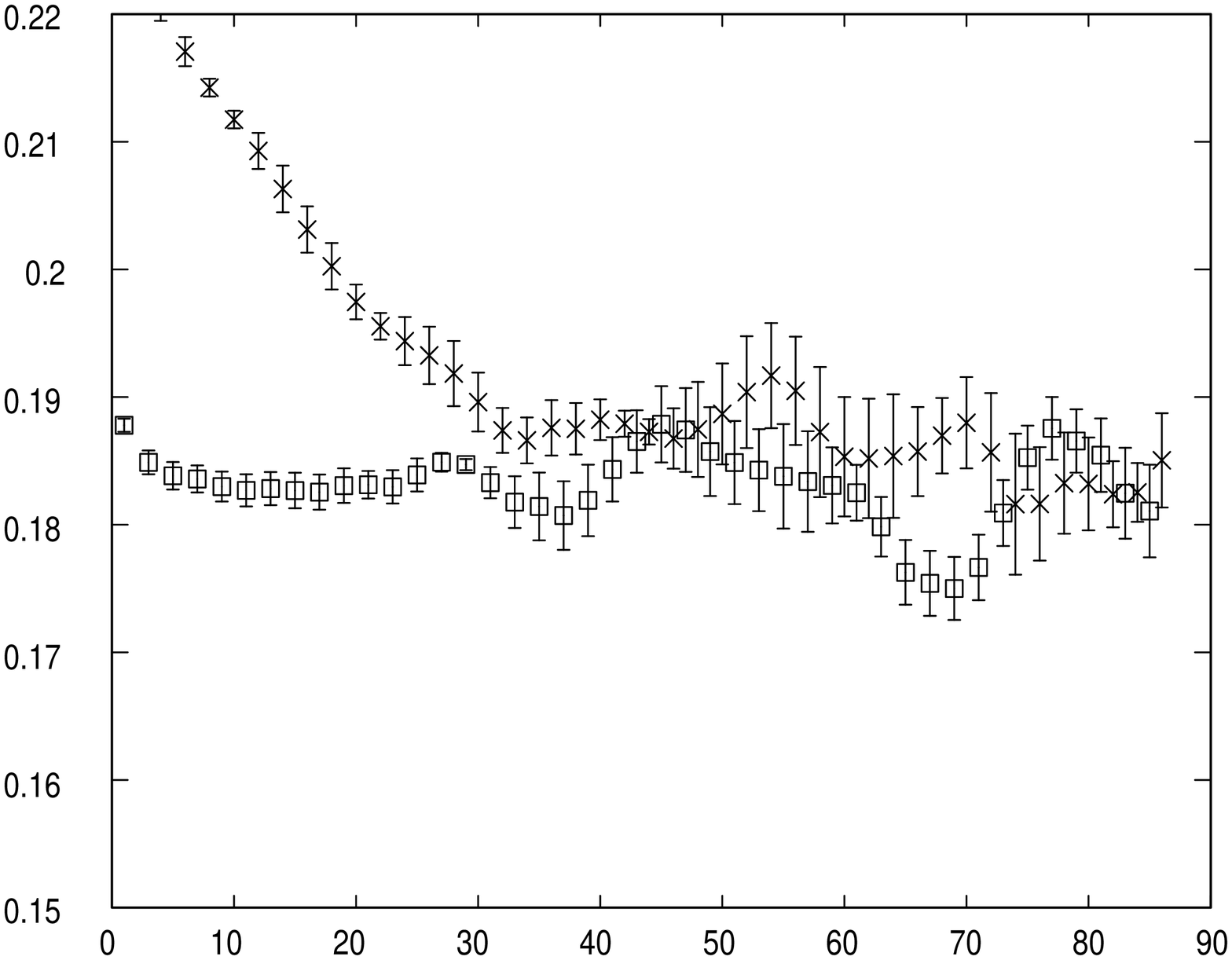}}}
\end{picture}
\caption{
$\theta$ vs. $t$ for the Ising model with $m_0$=0.04.
}
\label{fig2b}
\end{figure}

\renewcommand{\thefigure}{2.c}

\begin{figure}[t]\centering 
\epsfysize=12cm
\epsfclipoff
\fboxsep=0pt
\setlength{\unitlength}{1cm}
\begin{picture}(13.6,12)(0,0)
\put( 0.7, 8.1){\makebox(0,0){$\theta$}}
\put( 9.8,  .2){\makebox(0,0){$t$}}
\put( 6.5, 7.0){\makebox(0,0){\bf\large $\times$\ \ Metropolis}}
\put( 6.5, 2.5){\makebox(0,0){\bf\large $\Box$\ \ Heat-bath, $m_0$ = 0.06,
   L=128}}
\put(0,0){{\epsffile{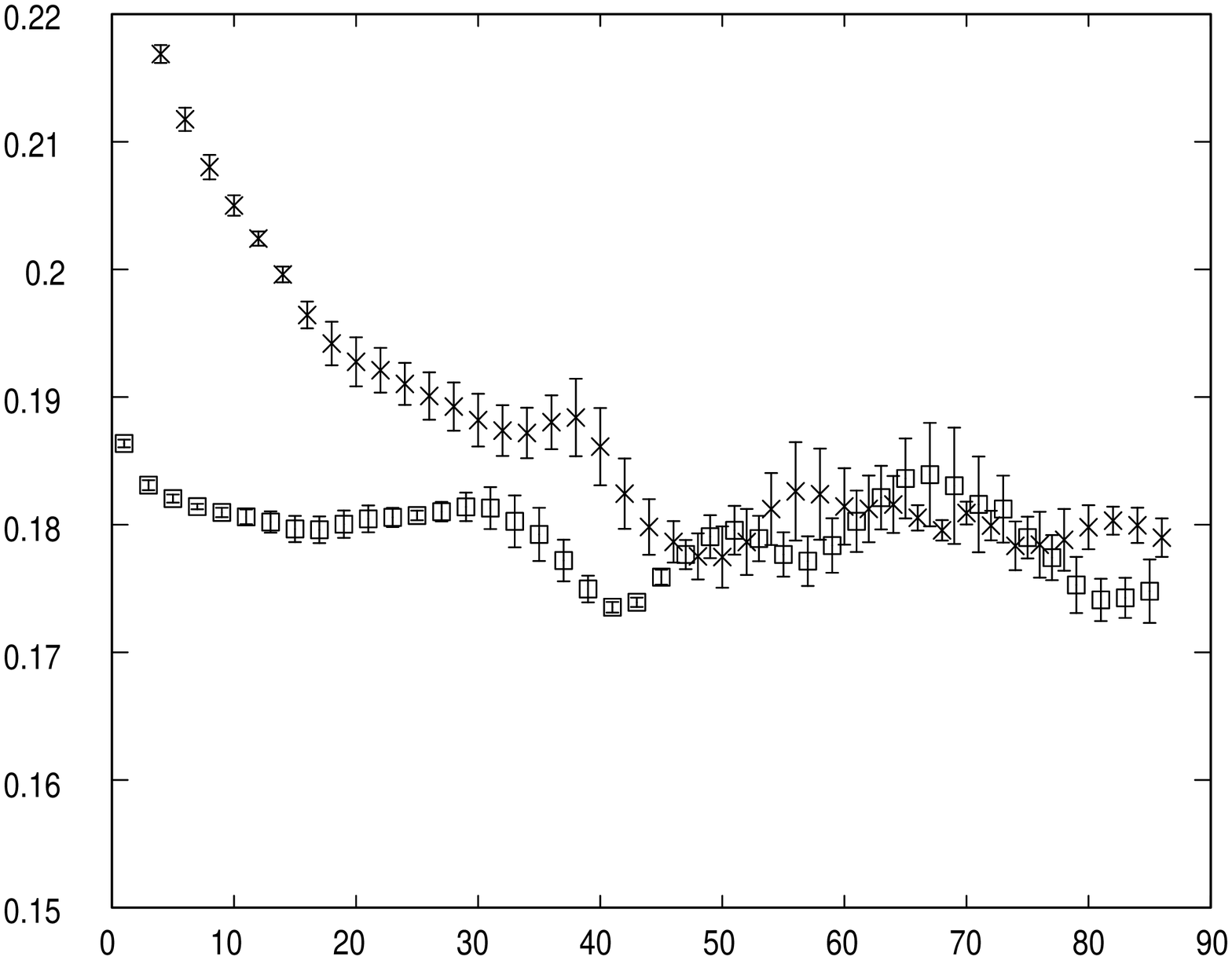}}}
\end{picture}
\caption{
$\theta$ vs. $t$ for the Ising model with $m_0$=0.06.
}
\label{fig2c}
\end{figure}

\renewcommand{\thefigure}{2.d}

\begin{figure}[b]\centering 
\epsfysize=12cm
\epsfclipoff
\fboxsep=0pt
\setlength{\unitlength}{1cm}
\begin{picture}(13.6,12)(0,0)
\put( 0.7, 8.1){\makebox(0,0){$\theta$}}
\put( 9.8,  .2){\makebox(0,0){$t$}}
\put( 6.5, 6.0){\makebox(0,0){\bf\large $\times$\ \ Metropolis}}
\put( 6.5, 1.5){\makebox(0,0){\bf\large $\Box$\ \ Heat-bath, $m_0$ = 0.08,
   L=128}}
\put(0,0){{\epsffile{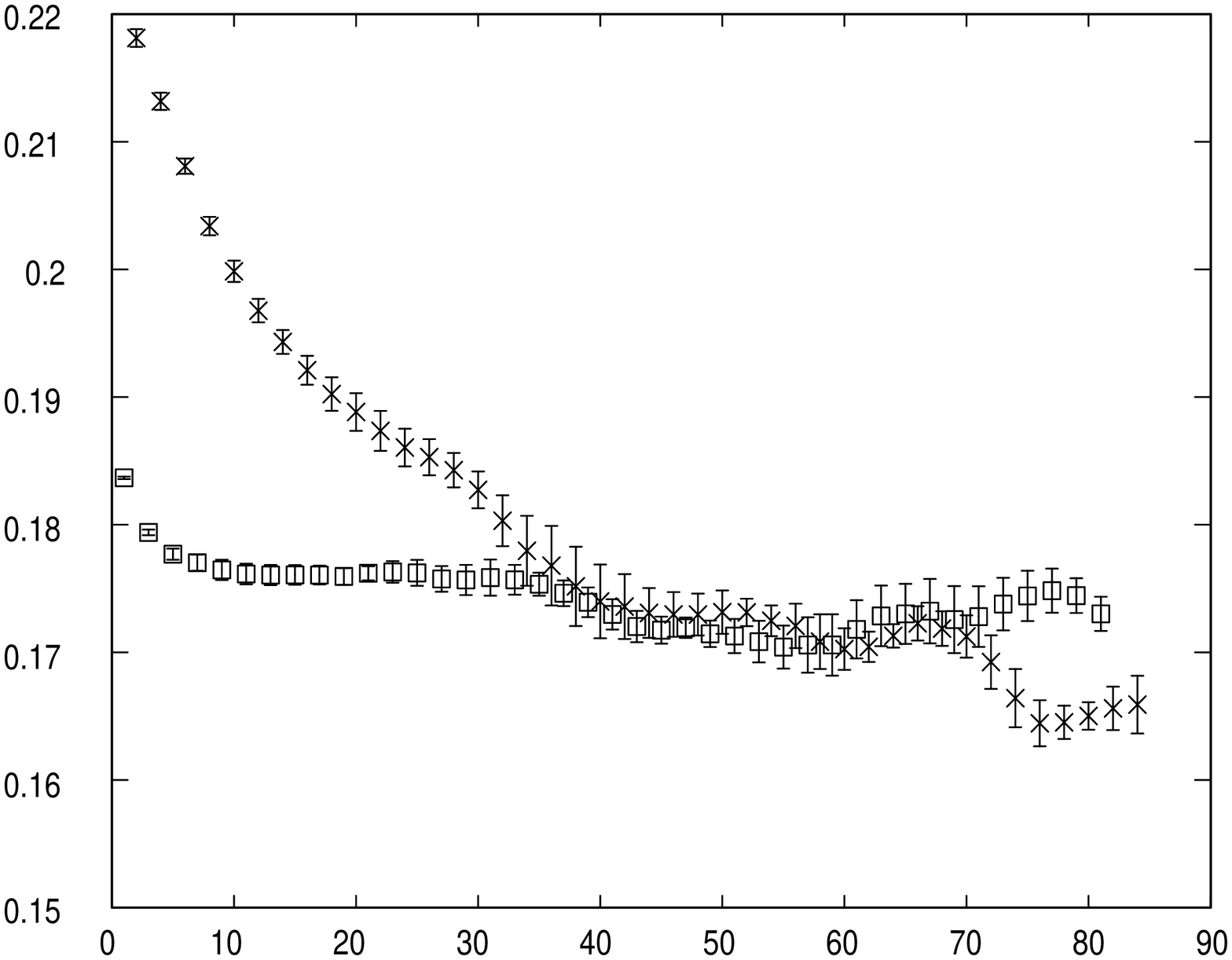}}}
\end{picture}
\caption{
$\theta$ vs. $t$ for the Ising model with $m_0$=0.08.
}
\label{fig2d}
\end{figure}

\renewcommand{\thefigure}{3}
\begin{figure}[t]\centering 
\epsfysize=12cm
\epsfclipoff
\fboxsep=0pt
\setlength{\unitlength}{1cm}
\begin{picture}(13.6,12)(0,0)
\put( 0.3, 8.25){\makebox(0,0){$\theta-{d}/{z}$}}
\put( 9.8,  .2){\makebox(0,0){$t$}}
\put( 4.5, 2.0){\makebox(0,0){\bf\large $\times$\ \ Metropolis\hspace{1.3cm}}}
\put( 4.5, 3.0){\makebox(0,0){\bf\large \hspace{0.3cm}$\Box$\ \  Heat-bath, L=256}}
\put(0,0){{\epsffile{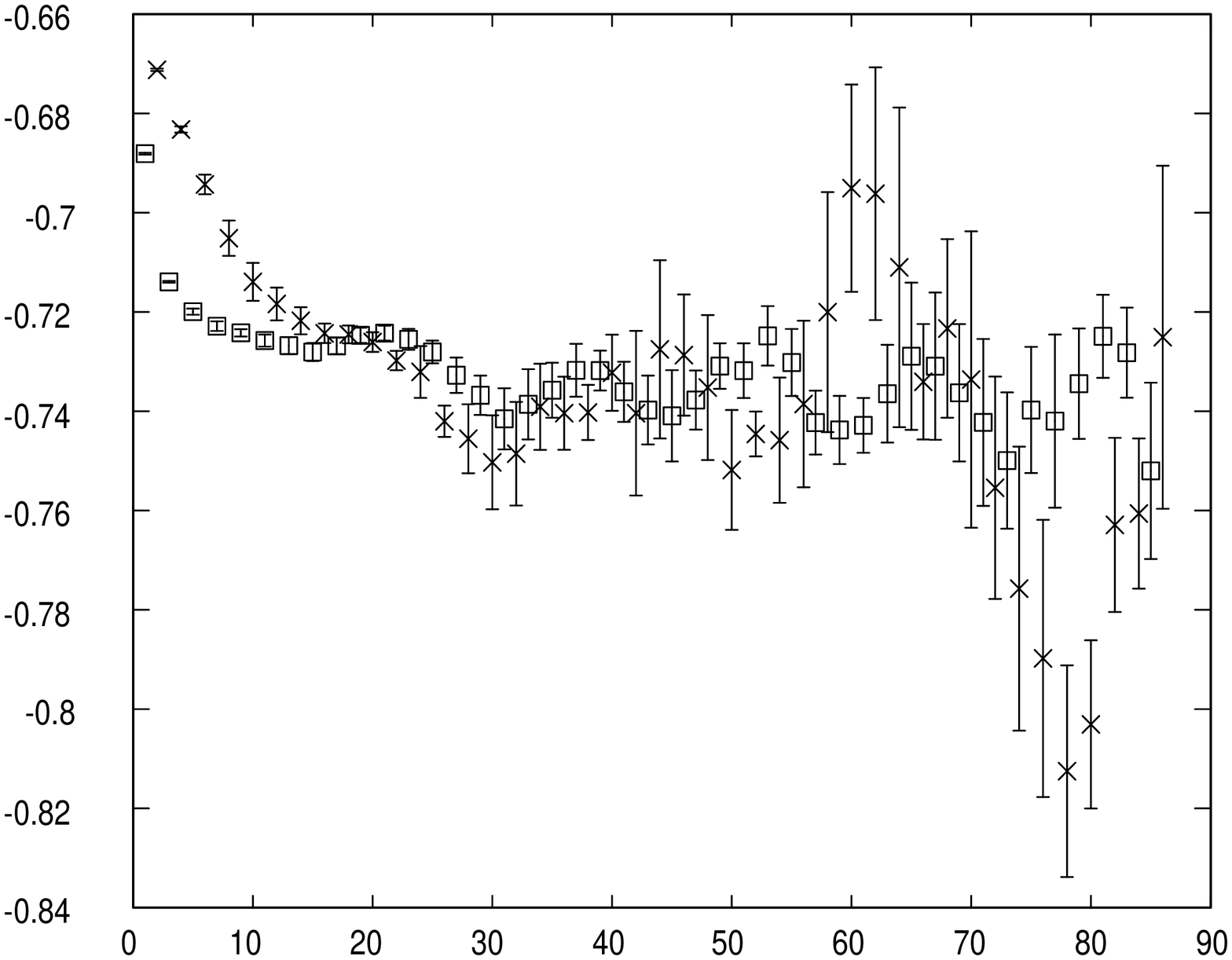}}}
\end{picture}
\caption{
$\theta-d/z$ vs. $t$ for the Ising model.
}
\label{fig3}
\end{figure}

\renewcommand{\thefigure}{4}
\begin{figure}[b]\centering 
\epsfysize=12cm
\epsfclipoff
\fboxsep=0pt
\setlength{\unitlength}{1cm}
\begin{picture}(13.6,12)(0,0)
\put( 0.3, 7.7){\makebox(0,0){$\frac{M(t)}{M(0)}$}}
\put( 9.8,  .2){\makebox(0,0){$t$}}
\put( 4.0, 7.0){\makebox(0,0){\bf\large Heat-bath, $m_0$=0.04}}
\put( 4.5, 3.0){\makebox(0,0){\bf\large Metropolis}}
\put(10.0, 8.0){\makebox(0,0){36}}
\put(10.6, 7.3){\makebox(0,0){18}}
\put( 9.2, 7.0){\makebox(0,0){72,144}}
\put( 8.5, 6.3){\makebox(0,0){L=9}}
\put(10.0, 3.7){\makebox(0,0){36}}
\put(10.0, 3.0){\makebox(0,0){18}}
\put(11.7, 3.9){\makebox(0,0){72}}
\put(11.7, 3.6){\makebox(0,0){144}}
\put( 8.5, 2.6){\makebox(0,0){L=9}}
\put(0,0){{\epsffile{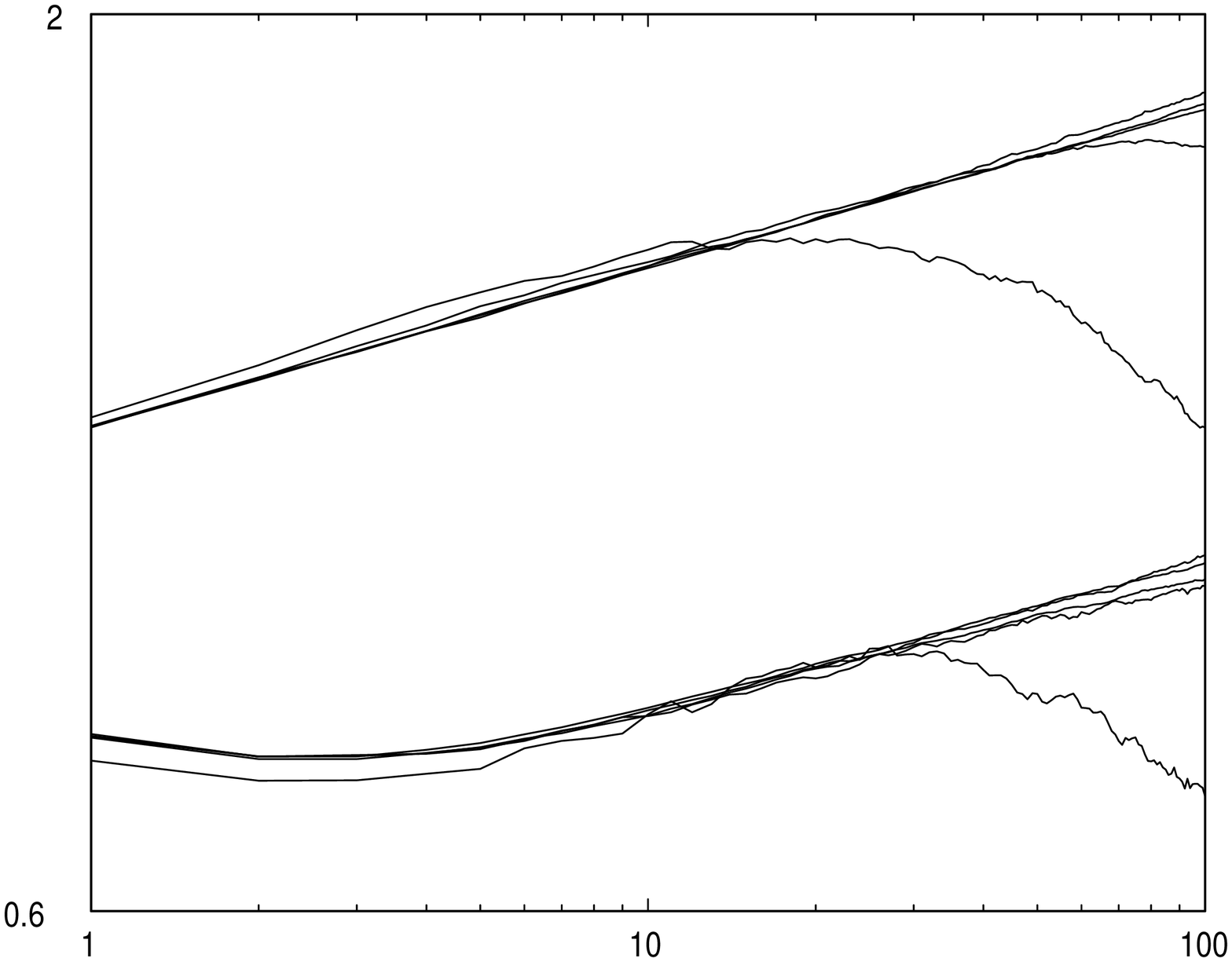}}}
\end{picture}
\caption{
The time evolution of the magnetization 
in double log scale 
for the Potts model with $m_0$=0.04
}
\label{fig4}
\end{figure}

\renewcommand{\thefigure}{5}
\begin{figure}[t]\centering 
\epsfysize=12cm
\epsfclipoff
\fboxsep=0pt
\setlength{\unitlength}{1cm}
\begin{picture}(13.6,12)(0,0)
\put( 0.5, 8.1){\makebox(0,0){$\theta$}}
\put( 9.8,  .2){\makebox(0,0){$t$}}
\put( 6.3, 8.1){\makebox(0,0){\bf\large $\Box$ Heat-bath, $m_0$=0.04, L=72}}
\put( 4.5, 4.5){\makebox(0,0){\bf\large $\times$ Metropolis}}
\put(0,0){{\epsffile{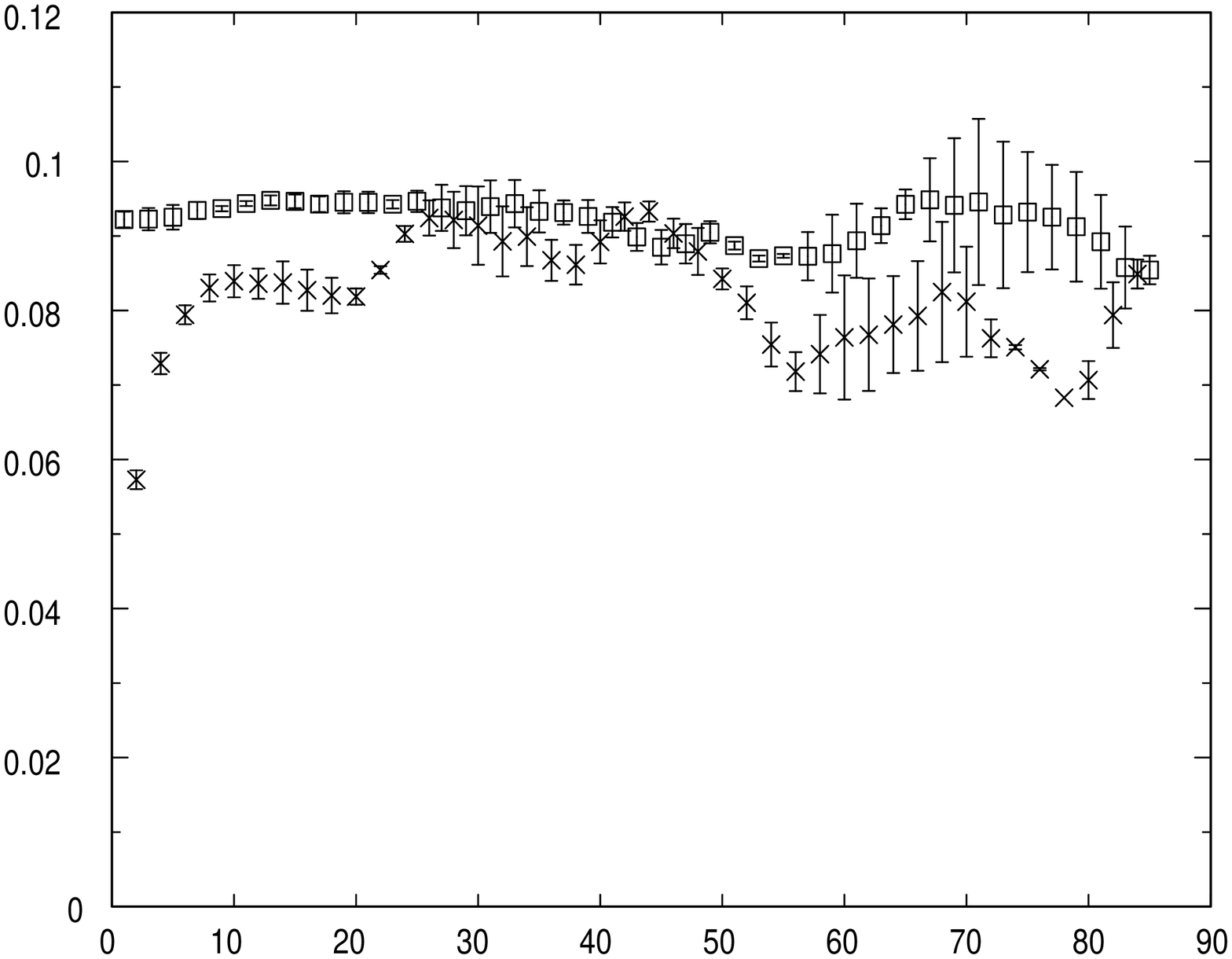}}}
\end{picture}
\caption{
$\theta$ vs. $t$
for the Potts model with $m_0$=0.04.
}
\label{fig5}
\end{figure}

\renewcommand{\thefigure}{6}
\begin{figure}[b]\centering 
\epsfysize=12cm
\epsfclipoff
\fboxsep=0pt
\setlength{\unitlength}{1cm}
\begin{picture}(13.6,12)(0,0)
\put( 0.5, 8.1){\makebox(0,0){$\theta$}}
\put( 9.8,  .2){\makebox(0,0){$t$}}
\put( 6.0, 3.1){\makebox(0,0){\bf\large Heat-bath, $m_0$=0.02}}
\put( 6.3, 2.2){\makebox(0,0){\bf\large $\Box$\ \ L=72}}
\put( 6.3, 1.6){\makebox(0,0){\bf\large \hspace{0.15cm}$\times$\ \ L=144}}
\put(0,0){{\epsffile{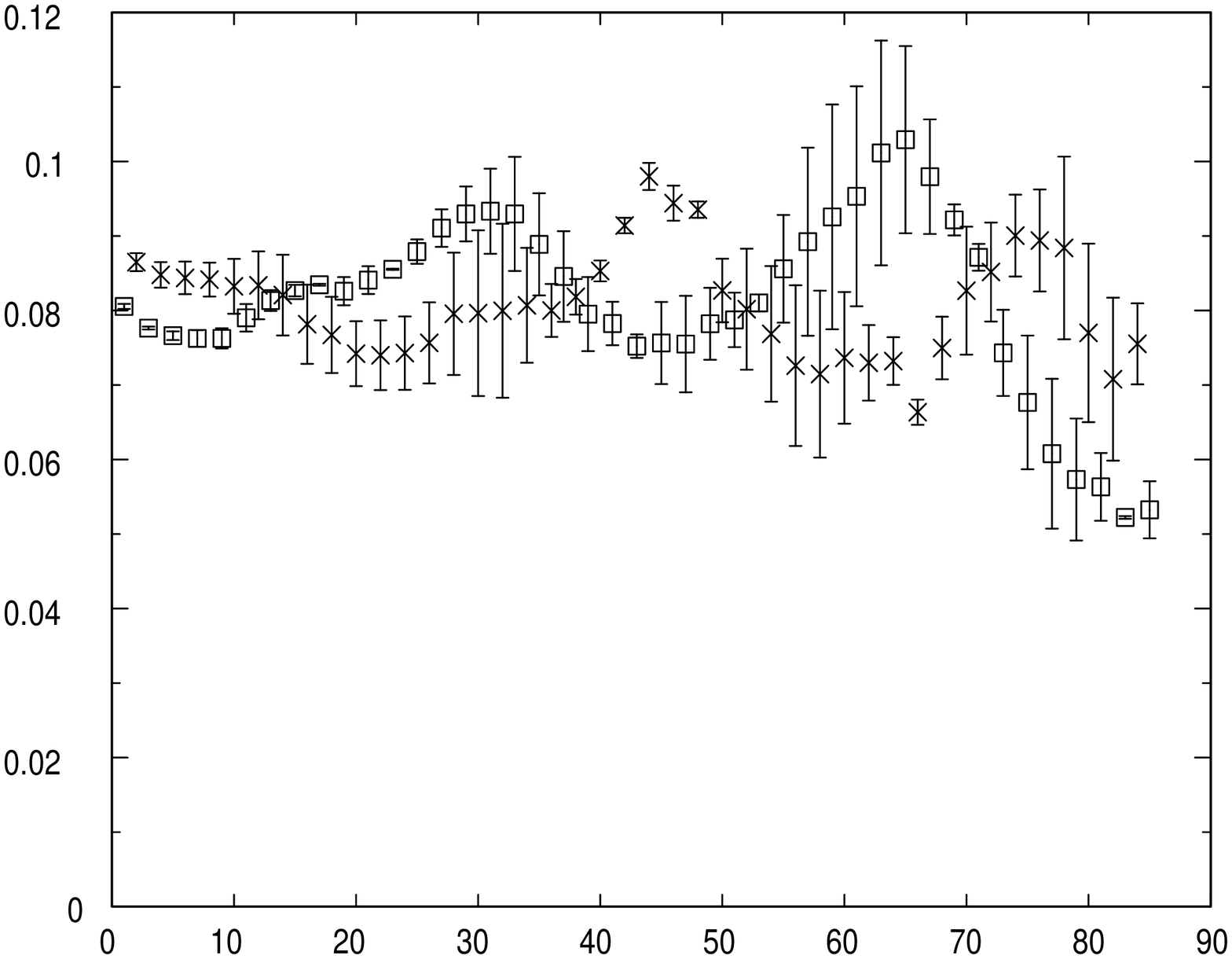}}}
\end{picture}
\caption{
$\theta$ vs. $t$
for the Potts model with $m_0$=0.02 and $L=72, 144$ for the Heat-bath
algorithm.
}
\label{fig6}
\end{figure}

\renewcommand{\thefigure}{7}
\begin{figure}[t]\centering 
\epsfysize=12cm
\epsfclipoff
\fboxsep=0pt
\setlength{\unitlength}{1cm}
\begin{picture}(13.6,12)(0,0)
\put( 0.7, 7.1){\makebox(0,0){$A(t)$}}
\put( 9.8,  .2){\makebox(0,0){$t$}}
\put( 8.0, 5.6){\makebox(0,0){\bf\large Heat-bath}}
\put( 6.0, 3.7){\makebox(0,0){\bf\large Metropolis}}
\put( 7.7, 2.1){\makebox(0,0){\bf\large L=72,144,288,576}}
\put(0,0){{\epsffile{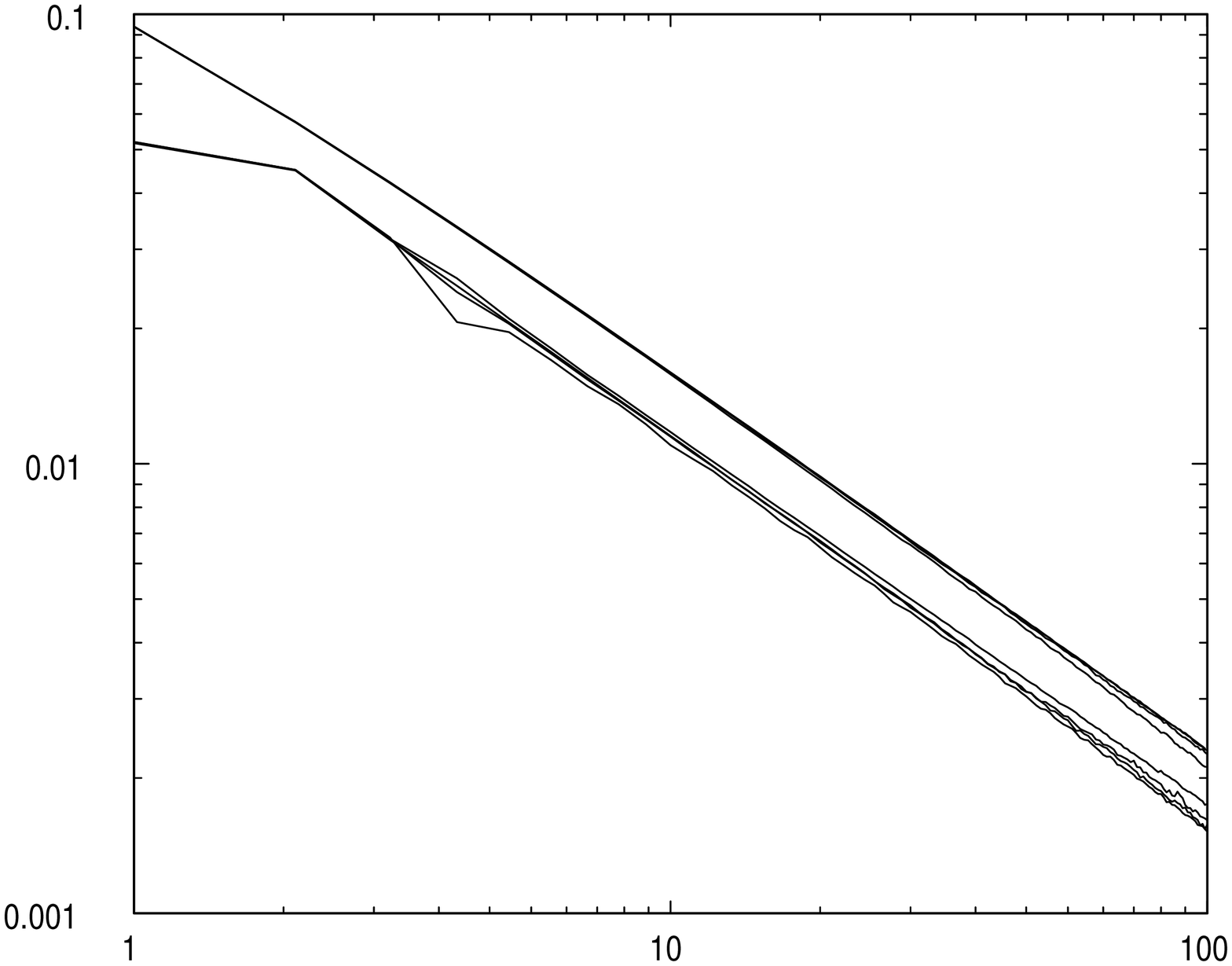}}}
\end{picture}
\caption{
The auto-correlation as a function of t in double log scale
for the Potts model.
}
\label{fig7}
\end{figure}

\renewcommand{\thefigure}{8}
\begin{figure}[b]\centering 
\epsfysize=12cm
\epsfclipoff
\fboxsep=0pt
\setlength{\unitlength}{1cm}
\begin{picture}(13.6,12)(0,0)
\put( 0.3, 7.5){\makebox(0,0){$\theta-{d}/{z}$}}
\put( 9.8,  .2){\makebox(0,0){$t$}}
\put( 4.5, 3.2){\makebox(0,0){\bf\large \hspace{0.5cm}$\Box$\ \ Heat-bath, L=288}}
\put( 4.0, 2.4){\makebox(0,0){\bf\large $\times$\ \ Metropolis}}
\put(0,0){{\epsffile{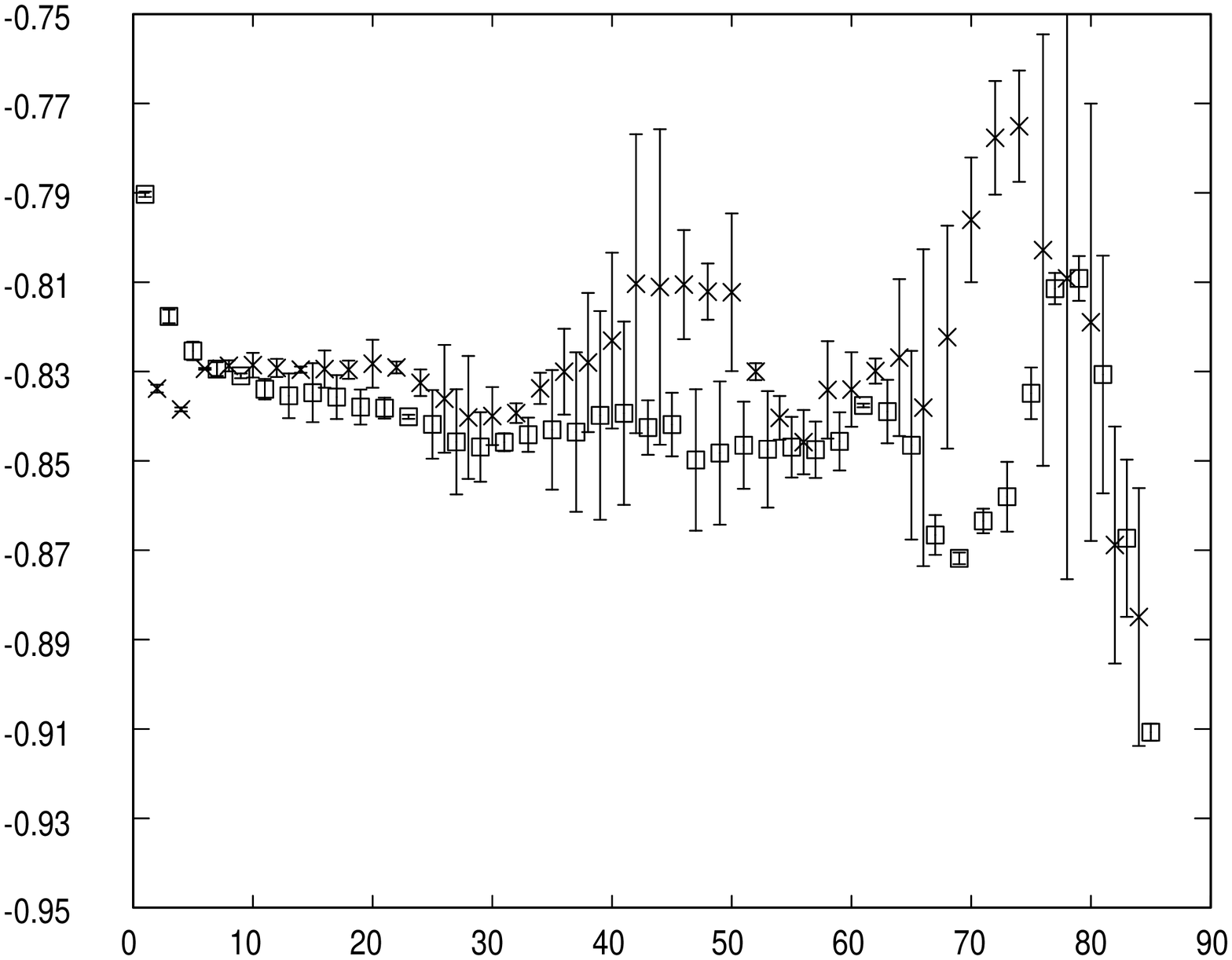}}}
\end{picture}
\caption{
$\theta-d/z$ vs. $t$ for the Potts model.
}
\label{fig8}
\end{figure}

\renewcommand{\thefigure}{9}
\begin{figure}[t]\centering 
\epsfysize=12cm
\epsfclipoff
\fboxsep=0pt
\setlength{\unitlength}{1cm}
\begin{picture}(13.6,12)(0,0)
\put( 0.2, 7.5){\makebox(0,0){$\frac{M^{(2)}(t)}{M^{(2)}(1)}$}}
\put( 9.8,  .2){\makebox(0,0){$t$}}
\put( 3.5, 4.7){\makebox(0,0){\bf\large \ \ Heat-bath}}
\put( 7.5, 3.0){\makebox(0,0){\bf\large \ \ Metropolis, L=9,...,288}}
\put(0,0){{\epsffile{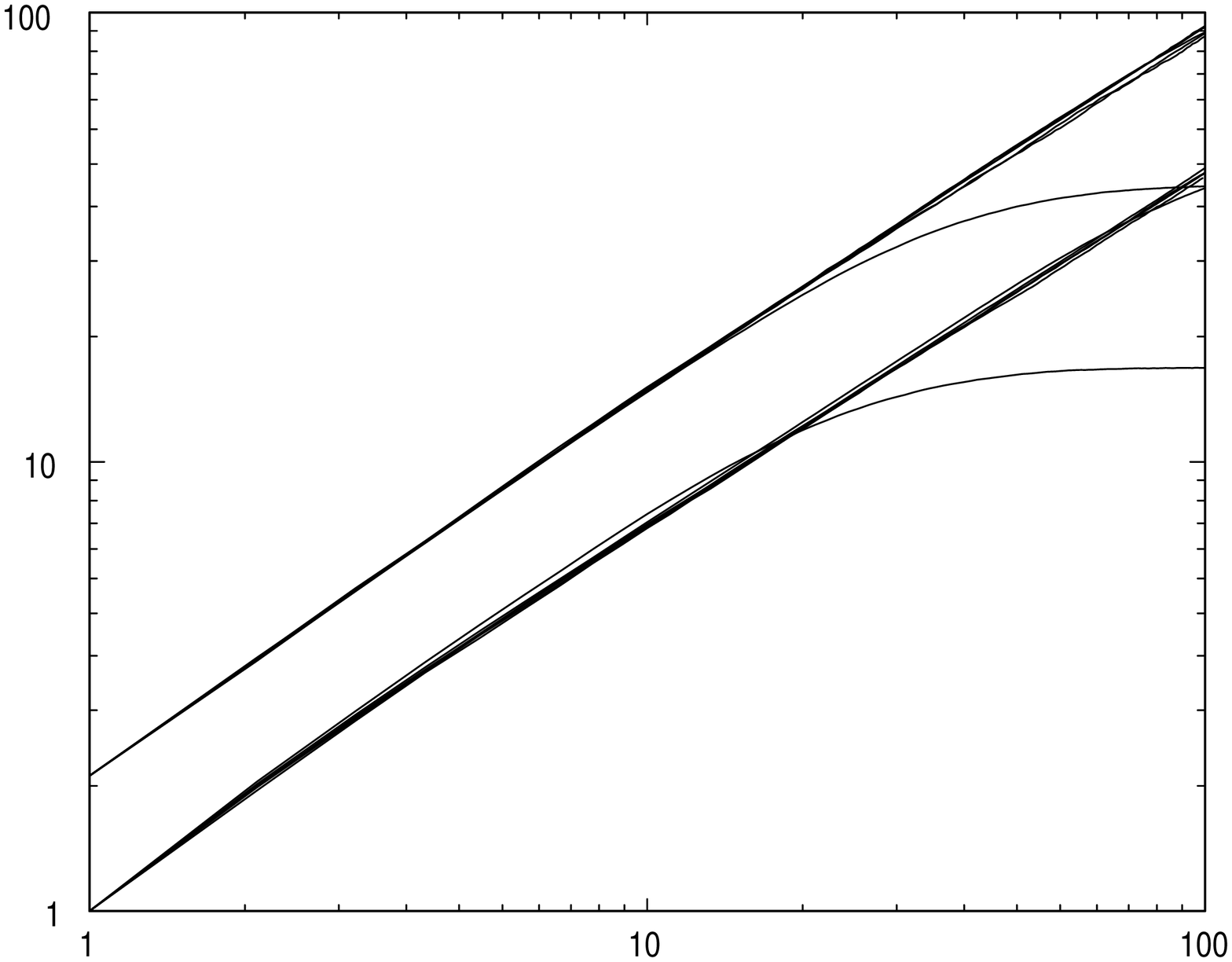}}}
\end{picture}
\caption{
The time evolution of the second moment in double log scale for 
the Potts model.
}
\label{fig9}
\end{figure}

\renewcommand{\thefigure}{10}
\begin{figure}[b]\centering 
\epsfysize=12cm
\epsfclipoff
\fboxsep=0pt
\setlength{\unitlength}{1cm}
\begin{picture}(13.6,12)(0,0)
\put( -0.3, 7.5){\makebox(0,0){$(d-{2\beta}/{\nu})/z$}}
\put( 9.8,  .2){\makebox(0,0){$t$}}
\put( 4.5, 7.0){\makebox(0,0){\bf\large $\times$\ \ Metropolis}}
\put( 3.5, 2.2){\makebox(0,0){\bf\large \hspace{3.2cm}$\Box$\ \ Heat-bath, L=288}}
\put(0,0){{\epsffile{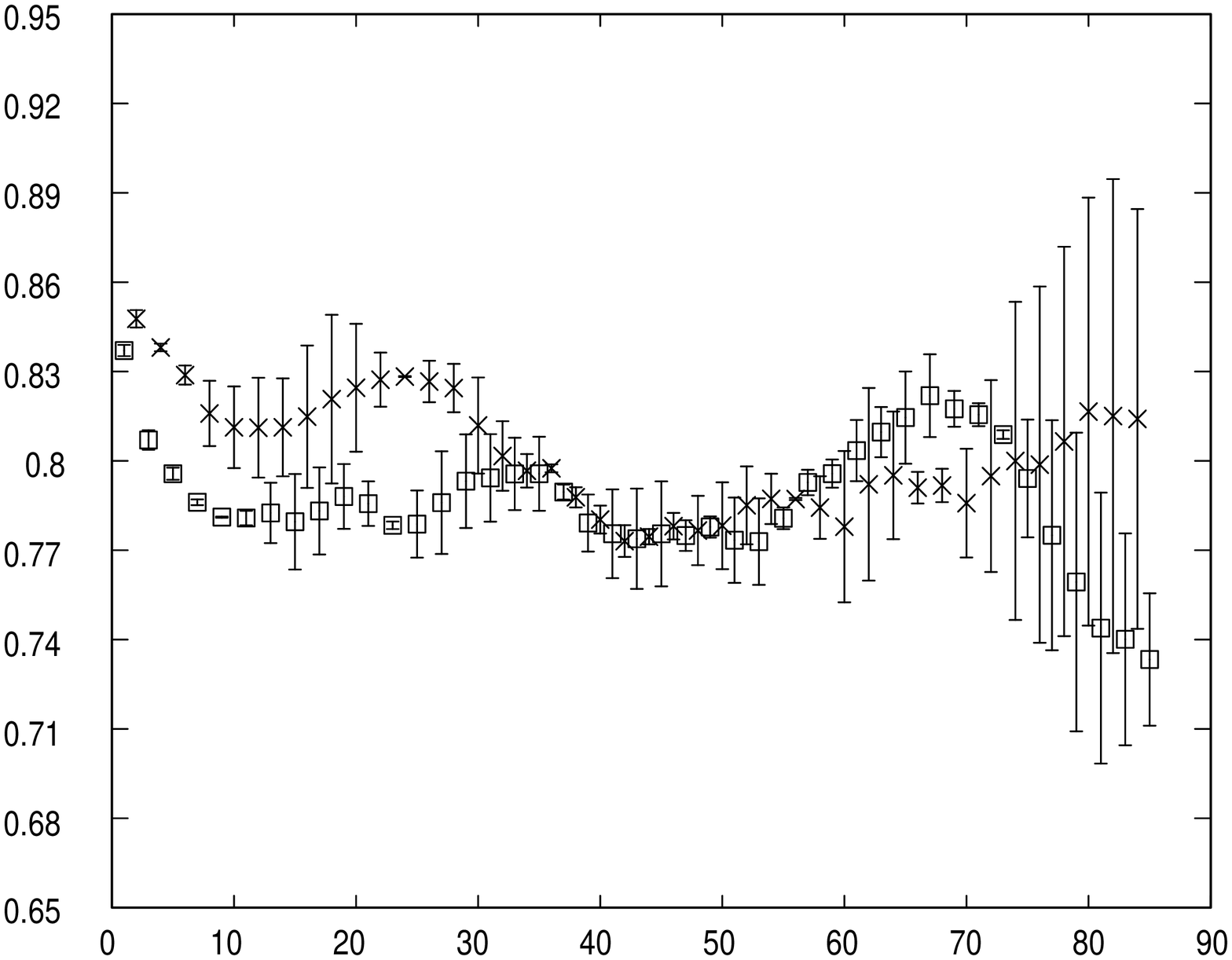}}}
\end{picture}
\caption{
$(d-2\beta/\nu)/z$ vs. $t$ for the Potts model. 
}
\label{fig10}
\end{figure}

\end{document}